\documentstyle[aps,preprint]{revtex}

\begin{document}
\title{STOCHASTIC RESONANCE IN A SYSTEM OF FERROMAGNETIC PARTICLES}

\author{ A. P\'{e}rez-Madrid, and J. M. Rub\'{\i} \\Departament de
F\'{\i}sica Fonamental\\Facultat de F\'{\i}sica\\ Universitat de
Barcelona\\Diagonal 647, 08028 Barcelona, Spain}

\maketitle
\parskip 2ex

\begin{abstract}

We show that a dispersion of  monodomain ferromagnetic particles in a
solid phase exhibits stochastic resonance when a driven linearly
polarized magnetic field is applied.  By using an adiabatic approach,
we calculate the power spectrum, the distribution of residence times
and the mean first passage time.The behavior of these quantities is
similar to their corresponding ones in other systems in which
stochastic resonance has also been observed.

\end{abstract}


\newpage

\section{Introduction}

The phenomenon known as stochastic resonance (SR) was first predicted
by Benzi et al. \cite{kn:Benzi}, and consists of the coherent
response of a multistable stochastic system to a driven periodic
signal.  Up to now,  SR has been observed in diverse physical
situations as in lasers, in electron paramagnetic resonance, or in
free standing magnetoelastic beams. The description of the phenomena
as well as its fundamentals and applications are included in the
recent reviews by F. Moss \cite{kn:Moss}, and K. Wiesenfeld and
F. Moss \cite{kn:Wiesenfeld} (see also in this context
refs.
\cite{kn:Jung} and
\cite{kn:nato}).

Our purpose in this paper is to show theoretical predictions about
the occurrence of the phenomenon in a system of ferromagnetic
monodomain particles dispersed in a solid phase (a crystalline
polymer, for example) when an alternating magnetic field is imposed.
In a ferromagnet the interdomain  walls are of the order $10^{-6}$
cm,  then particles whose size is  of this order of magnitude or less
may be considered monodomains \cite{kn:LanMir}. Such a particles are
always magnetized to the spontaneous magnetization $M_{s}$. For these
fine ferromagnetic particles the energy consists of contributions
coming from two competing mechanisms which tend to orient the
magnetic moment: potential energy due to the field and energy of
anisotropy. The energy is therefore a nonlinear function of the
orientation angle, which is precisely the stochastic variable.  This
fact was already taken into account by N\'{e}el \cite{kn:Neel} to
estimate the relaxation time for the magnetic moment in magnetic
powders. As we will show, under certain conditions the ferromagnetic
particle constitutes a bistable stochastic system, with the external
field providing a periodic contribution.

The stochastic behavior of systems of ferromagnetic particles was
discussed in ref. \cite{kn:Brown} where a Fokker-Planck equation for
the probability density of the orientations of the particles was
derived. This equation is related to the Landau-Gilbert equation
\cite{kn:Landau}, \cite{kn:Gilbert} in which a stochastic source
accounting for Brownian motion of the magnetic moment is added. The
procedure is, however, restricted to the case in which the external
magnetic field is constant in time. This theory provides the
framework for our subsequent analysis and consequently must be
extended to the case of a time-dependent magnetic field.

We have distributed the paper in the following way. In section II we
introduce our model and from the Gilbert-Landau equation  we
establish the kinetic equation for the probability distribution of
the magnetic moment. In section III, by using an adiabatic
approximation, we compute the power spectrum of the fluctuations of
the magnetic moment, whereas, in section IV, and within the framework
of the same approach,  we calculate the probability distribution of
residence times and the mean first passage time. Finally, in section
V we give numerical values of the characteristic parameters of our
system and discuss our main results. Additionally, we show that due
to the very short time scale that rules the relaxation of the system,
the adiabatic approach is justified.

\section{The dispersion of ferromagnetic particles}

We consider an assembly of single-domain ferromagnetic particles
dispersed in a solid phase at a concentration which is assumed to be
low enough to avoid magnetic interactions among them. When we apply
an external uniform a.c. magnetic field $\vec{H}(t)= \vec{H}_{0}
sin\omega_{0}t$, $\vec{H}_{0}$ being the magnetic field strength and
$\omega_{0}$ its angular frequency, the energy of each particle
splits up into contributions coming from the external field and the
crystalline anisotropy \cite{kn:LanMir} and is given by

\begin{equation}\label{eq:a1}
U(t) = - \vec{m}\cdot\vec{H}(t) + K_{a}V_{p}(\hat{m}\cdot\hat{s})^2
\;\; .
\end{equation}

\noindent Here $\vec{m}=m_{s}\hat{m}$ is the magnetic dipole moment,
$m_{s}=M_{s}V_{p}$ the magnetic moment strength, with $M_{s}$ being
the saturation magnetization and $V_{p}$ the volume of the particle,
$K_{a}>0$ is the anisotropy constant and $\hat{s}$ is a unit vector
perpendicular to the symmetry axis of the particle.

The dynamics of the magnetic moment $\vec{m}$ is governed by the
Gilbert equation \cite{kn:Gilbert}, \cite{kn:Brownbook}

\begin{equation}\label{eq:a2}
\frac{1}{\gamma}\frac{d\vec{m}}{dt} = \vec{m}\times(\vec{H}_{eff} +
\vec{H}_{d}) \;\; ,
\end{equation}

\noindent where $\gamma(=-e/mc)$ is the gyromagnetic ratio. From
(\ref{eq:a2}) one may identify the two mechanisms responsible for the
variation of $\vec{m}$. The effective field

\begin{equation}\label{eq:a3}
\vec{H}_{eff} \equiv \frac{\partial U}{\partial \vec{m}} = \vec{H}(t)
- 2\frac{K_{a}V_{p}}{m_{s}}(\hat{m}\cdot\hat{s})\hat{s}
\end{equation}

\noindent which implies a Larmor precessional motion of $\vec{m}$ and
the mean field

\begin{equation}\label{eq:a4}
\vec{H}_{d} \equiv -\eta \frac{d\vec{m}}{dt}
\end{equation}

\noindent that introduces a damping whose microscopic origin lies in
the collisions among the electrons participating in the formation of
the magnetic moment of the domain. In eq. (\ref{eq:a4}) $\eta$ is a
damping coefficient.

Eq. (\ref{eq:a2}) can be solved self-consistently giving

\begin{equation}\label{eq:a5}
\frac{d\vec{m}}{dt} = \vec{\omega}_{L}\times\vec{m} +
h\vec{m}\times\vec{H}_{eff}\times\vec{m}\;\; ,
\end{equation}

\noindent where $\vec{\omega}_{L}=-m_{s}g\vec{H}_{eff}$ is the Larmor
angular frequency of the precessional motion executed by the magnetic
moment of a dipole. Moreover, we have introduced the quantities

\begin{equation}\label{eq:a6}
g = \frac{\gamma}{m_{s}(1+{\eta}^2m_{s}^2{\gamma}^2)}\;\; ,
\end{equation}

\noindent and

\begin{equation}\label{eq:a7}
h = -\frac{\eta{\gamma}^2}{(1+{\eta}^2m_{s}^2{\gamma}^2)}\;\; .
\end{equation}

\noindent When the external field is constant, the precessional
motion is extinguished in a time scale
$\tau_{0}=(m_{s}hH_{eff})^{-1}$, obtained by comparison of the left
hand term and the second right hand term in (\ref{eq:a5}). Thus when
Brownian motion is absent, $\vec{m}$ become parallel to
$\vec{H}_{eff}$ for times larger than $\tau_{0}$.  Keeping only first
order terms in the damping coefficient, one obtains the Landau
equation \cite{kn:Landau}

\begin{equation}\label{eq:a8}
\frac{d\vec{m}}{dt} = -\gamma \vec{H}_{eff}\times\vec{m} +
\lambda\vec{m}\times\vec{H}_{eff}\times\vec{m}\;\; ,
\end{equation}

\noindent where $\lambda$ may be
identified as $\eta{\gamma}^2$.

The presence of thermal noise was considered by Brown \cite{kn:Brown}
by simply adding the random field $\vec{H}_{r}$ to the Gilbert
equation (\ref{eq:a2}). This equation thus becomes a nonlinear
Langevin equation with multiplicative noise

\begin{equation}\label{eq:a9}
\frac{1}{\gamma}\frac{d\vec{m}}{dt} = \vec{m}\times(\vec{H}_{eff} +
\vec{H}_{d} + \vec{H}_{r})\;\; .
\end{equation}

\noindent The random term constitutes a gaussian stochastic process
with zero mean and a fluctuation-dissipation theorem given by

\begin{equation}\label{eq:h1}
\langle \vec{H}_{r}(t') \vec{H}_{r}(t'+t)\rangle = 2K_{B}T\eta
\vec{\vec{1}}\delta(t) \; ,
\end{equation}

\noindent where $\vec{\vec{1}}$ is the unit tensor.

Following the standard procedure, it is possible to derive the
Fokker-Planck equation related to (\ref{eq:a9}). One then obtains
\cite{kn:Brown}, \cite{kn:Rubi}

\begin{equation}\label{eq:a10}
\frac{\partial \psi}{\partial t} - \vec{S}\cdot\vec{\omega}_{L}\psi =
-\frac{1}{2\tau}\vec{S}\cdot\psi\vec{S}\left(\frac{U(t)}{K_{B}T} +
log \psi\right)\;\; ,
\end{equation}

\noindent where $\psi(\hat{m},t)$ is the distribution function for
the orientations of the vector $\hat{m}$ and
$\vec{S}=\hat{m}\times\frac{\partial}{\partial\hat{m}}$ is the
rotational operator. From this equation we infer the appearance of
the relaxation time $\tau=(-2K_{B}Th)^{-1}$, corresponding to the
time scale in which one achieves the stationary state where the
probability fluxe is constant.

If the external magnetic field is applied along the direction of the
easy axis of magnetization, the problem posed by eq. (\ref{eq:a10})
has axial symmetry. In this case, the energy of the system can be
written as

\begin{equation}\label{eq:a11}
U = -m_{s}H_{0} sin\omega_{0}t cos\theta + K_{a}V_{p} sin^2\theta\;\;
\end{equation}

TEX FILE ARRIVED TRUNCATED. SUBMITTER DIDN'T CARE, NEITHER DO WE.

\end{document}